\documentclass[aps,prl,reprint,floatfix,citeautoscript,longbibliography,superscriptaddress]{revtex4-1}
\usepackage{graphicx}
\usepackage{ulem}
\usepackage{bm,amsmath,amssymb,mathrsfs,dcolumn}
\usepackage[colorlinks=true, linkcolor=blue, citecolor=blue, urlcolor=blue, linktoc=page, bookmarks=false, pdfstartview={FitH}, pdfborder={0 0 0.0 [3 3]}]{hyperref} 
\usepackage[usenames]{xcolor}
\hypersetup{pdfborder=0 0 0,colorlinks=true,citecolor=blue,linkcolor=blue}

\begin{document}

\title{Mode-Dependent Damping in Metallic Antiferromagnets Due to Inter-Sublattice Spin Pumping}
\author{Qian Liu}
\altaffiliation{These authors contributed equally to this work.}
\affiliation{The Center for Advanced Quantum Studies and Department of Physics, Beijing Normal University, 100875 Beijing, China}
\author{H. Y. Yuan}
\altaffiliation{These authors contributed equally to this work.}
\affiliation{Department of Physics, South University of Science and Technology of China,
Shenzhen, Guangdong, 518055, China}
\author{Ke Xia}
\affiliation{The Center for Advanced Quantum Studies and Department of Physics, Beijing Normal University, 100875 Beijing, China}
\affiliation{Synergetic Innovation Center for Quantum Effects and Applications (SICQEA), Hunan Normal University, Changsha 410081, China}
\author{Zhe Yuan}
\email[Corresponding author: ]{zyuan@bnu.edu.cn}
\affiliation{The Center for Advanced Quantum Studies and Department of Physics, Beijing Normal University, 100875 Beijing, China}
\date{\today}
\begin{abstract}
Damping in magnetization dynamics characterizes the dissipation of magnetic energy and is essential for improving the performance of spintronics-based devices. While the damping of ferromagnets has been well studied and can be artificially controlled in practice, the damping parameters of antiferromagnetic materials are nevertheless little known for their physical mechanisms or numerical values. Here we calculate the damping parameters in antiferromagnetic dynamics using the generalized scattering theory of magnetization dissipation combined with the first-principles transport computation. For the PtMn, IrMn, PdMn and FeMn metallic antiferromagnets, the damping coefficient associated with the motion of magnetization ($\alpha_m$) is one to three orders of magnitude larger than the other damping coefficient associated with the variation of the N{\'e}el order ($\alpha_n$), in sharp contrast to the assumptions made in the literature. 
\end{abstract}
\maketitle

Damping describes the process of energy dissipation in dynamics and determines the time scale for a nonequilibrium system relaxing back to its equilibrium state. For magnetization dynamics of ferromagnets (FMs), the damping is characterized by a phenomenological dissipative torque exerted on the precessing magnetization \cite{Gilbert2004}. The magnitude of this torque that depends on material, temperature and magnetic configurations, has been well studied in experiment \cite{Heinrich1966,Bhagat1974,Heinrich1979,Mizukami2001The,Mizukami2001Ferromagnetic,Ingvarsson2002,Lubitz2003,Yakata2006,Weindler2014} and theory \cite{Kambersky1976,*Kambersky:prb07,Gilmore2007,*Gilmore2010,Starikov:prl10,Ebert:prl11,Liu2011,Tang2017}.

Recently, magnetization dynamics of antiferromagnets (AFMs)  \cite{Kimel2004,MacDonald2011,Marti2014,Jungwirth2016}, especially that controlled by an electric or spin current \cite{Nunez2006,Haney2007,Xu2008,Haney2008,Swaving2011,Hals2011,Zelezny2014,Qu2015,Zhang2016,Joseph2016,Fukami2016,Wadley2016}, has attracted lots of attention in the process of searching the high-performance spintronic devices. However, the understanding of AFM dynamics, in particular the damping mechanism and magnitude in real materials, is quite limited. Magnetization dynamics of a collinear AFM can be described by two coupled Landau-Lifshitz-Gilbert (LLG) equations corresponding to the precessional motion of the two sublattices, respectively \cite{Keffer1952}, i.e. ($i=1,2$)
\begin{eqnarray}
\dot{\mathbf m}_i=-\gamma \mathbf m_i\times\mathbf h_i+\alpha_i\mathbf m_i\times\dot{\mathbf m}_i,\label{eq:coupledm}
\end{eqnarray}
where $\gamma$ is the gyromagnetic ratio, $\mathbf{m}_i$ is the magnetization direction on the $i$-th sublattice and $\dot{\mathbf m}_i=\partial_t \mathbf m_i$. $\mathbf h_i$ is the effective magnetic field on $\mathbf m_i$, which contains the anisotropy field, the external field and the exchange field arising from the magnetization on the both sublattices. The last contribution to $\mathbf h_i$ makes the dynamic equation of one sublattice coupled to the equation of the other one. Specifically, if the free energy of the AFM is given by the following form $\mathcal F[\mathbf m_1,\mathbf m_2]\equiv\mu_0 M_s V\mathcal E[\mathbf m_1,\mathbf m_2]$ with the permeability of vacuum $\mu_0$, the magnetization on each sublattice $M_s$ and the volume of the AFM $V$, one has $\mathbf h_i=-\delta \mathcal E/\delta \mathbf m_i$. $\alpha_i$ in Eq.~\eqref{eq:coupledm} is the damping parameter representing the dissipation rate of the magnetization $\mathbf m_i$. Due to the sublattice permutation symmetry, the damping magnitudes of the two sublattices should be equal. This approach has been used to investigate the AFM resonance \cite{Keffer1952,Ross2015}, temperature gradient induced domain wall (DW) motion \cite{Selzer2016} and spin-transfer torques in an AFM$|$FM bilayer \cite{Helen2010}. 

An alternative way to deal with the AFM dynamics is introducing the net magnetization $\mathbf{m}\equiv\mathbf m_1+\mathbf m_2$ and the N\'{e}el order $\mathbf{n}\equiv\mathbf m_1-\mathbf m_2$ so that the precessional motion of $\mathbf{m}$ and $\mathbf{n}$ can be derived from the Lagrangian equation \cite{Hals2011}. The damping effect is then included artificially with two parameters $\alpha_m$ and $\alpha_n$ that characterize the dissipation rate of $\mathbf{m}$ and $\mathbf{n}$, respectively. This approach is widely used to investigate spin superfluid in an AFM insulator \cite{Halperin1969,Takei2014}, AFM nano-oscillator \cite{Cheng2016}, and DW motion induced by an electrical current \cite{Hals2011,Tveten2013}, spin waves \cite{Tveten2014} and spin-orbit torques \cite{Gomonay2016,Shiino2016}. 
Using the above definitions of $\mathbf m$ and $\mathbf n$, one can reformulate Eq.~\eqref{eq:coupledm} and derive the following dynamic equations
\begin{eqnarray}
\dot{\mathbf n}&=&\left(\gamma \mathbf h_m-\alpha_m\dot{\mathbf m}\right)\times\mathbf n+\left(\gamma\mathbf h_n-\alpha_n\dot{\mathbf n}\right)\times\mathbf m,\label{eq:norder}\\
\dot{\mathbf m}&=&\left(\gamma \mathbf h_m-\alpha_m\dot{\mathbf m}\right)\times\mathbf m+\left(\gamma \mathbf h_n-\alpha_n\dot{\mathbf n}\right)\times\mathbf n,\label{eq:morder}
\end{eqnarray}
where $\mathbf h_n$ and $\mathbf h_m$ are the effective magnetic fields exerted on $\mathbf n$ and $\mathbf m$, respectively. They can also be written as the functional derivative of the free energy \cite{Hals2011,Tveten2014}, i.e. $\mathbf h_n=-\delta \mathcal E/\delta \mathbf n$ and $\mathbf h_m=-\delta \mathcal E/\delta \mathbf m$. The damping parameters in Eqs.~(\ref{eq:coupledm}--\ref{eq:morder}) have the relation $\alpha_n = \alpha_m = \alpha_1/2=\alpha_2/2$ \cite{Helen2010}. Indeed, the assumption $\alpha_m = \alpha_n$ is commonly adopted in the theoretical study of AFM dynamics with only a few exceptions, where $\alpha_m$ is ignored in the current-induced skyrmion motion in AFM materials \cite{Velkov2016} and the magnon-driven DW motion \cite{Kim2014}. However, the underlying damping mechanism of an AFM and the relation between $\alpha_m$ and $\alpha_n$ have not been fully justified yet \cite{Gomonay2014,Atxitia2017}.

In this paper, we generalize the scattering theory of magnetization dissipation in FMs \cite{Brataas2008,*Brataas2011} to AFMs and calculate the damping parameters from first-principles for metallic AFMs PtMn, IrMn, PdMn and FeMn. The damping coefficients in an AFM are found to be strongly mode-dependent with $\alpha_m$ up to three orders of magnitude larger than $\alpha_n$. By analyzing the dependence of damping on the disorder and spin-orbit coupling (SOC), we demonstrate that $\alpha_n$ arises from SOC in analog to the Gilbert damping in FMs, while $\alpha_m$ is dominated by the spin pumping effect between sublattices.


{\it\color{red}Theory.---}In analogue to the scattering theory of magnetization dissipation in FMs \cite{Brataas2008,*Brataas2011}, the damping parameters in AFMs, $\alpha_n$ and $\alpha_m$, can be expressed in terms of the scattering matrix. Following the previous definition of the free energy, the energy dissipation rate of an AFM reads
\begin{eqnarray}
\dot E &=& -\mu_0 M_s V \dot{\mathcal E}=\mu_0 M_s V \left(-\frac{\delta \mathcal E}{\delta\mathbf m}\cdot \dot{\mathbf m}-\frac{\delta \mathcal E}{\delta\mathbf n}\cdot \dot{\mathbf n}\right)\nonumber\\
&=&\mu_0 M_s V(\mathbf h_m \cdot \dot{\mathbf m} + \mathbf h_n \cdot \dot{\mathbf n}).
\end{eqnarray}
By replacing the effective fields $\mathbf h_m$ and $\mathbf h_n$ by the time derivative of magnetization order and N\'{e}el order using Eq. (\ref{eq:norder}) and (\ref{eq:morder}), one arrives at \cite{SM}
\begin{equation}
\dot E = \frac{\mu_0 M_s V}{\gamma}\left(\alpha_n\dot{\mathbf n}^2+\alpha_m\dot{\mathbf m}^2\right). \label{eq:edot}
\end{equation}

\begin{figure}[t]
\includegraphics[width=0.8\columnwidth]{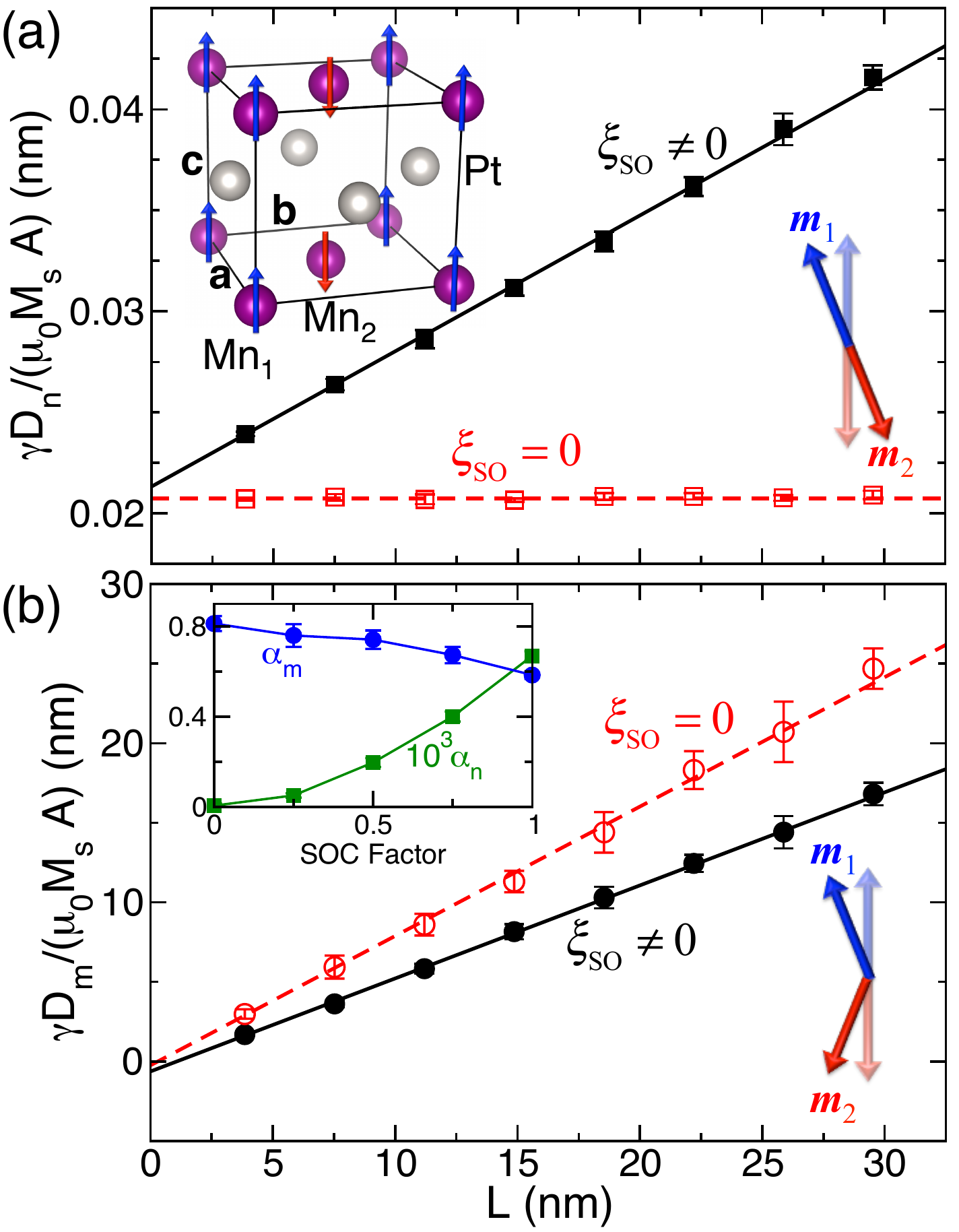}
\caption{Calculated energy dissipation rate as a function of the length of PtMn due to variation of the order parameters $\mathbf n$ (a) and $\mathbf m$ (b). $A$ is the cross-sectional area of the lateral supercell. Arrows in each panels illustrate the dynamical modes of the order parameters. The empty symbols are calculated without spin-orbit interaction. The inset of panel (a) shows atomic structure of PtMn with collinear AFM order. The inset in (b) shows calculated $\alpha_n$ and $\alpha_m$ as a function of the scaled SOC strength. The factor 1 corresponds to the real SOC strength that is determined by the derivative of the self-consistent potentials.}\label{fig:1}
\end{figure}

If we place an AFM between two semi-infinite nonmagnetic metals, the propagating electronic states coming from the metallic leads are partly reflected and transmitted. The probability amplitudes of the  reflection and transmission form the so-called scattering matrix $\mathbf S$ \cite{Datta1995}. For such a scattering structure with only the order parameter $\mathbf n$ of the AFM varying in time (see the insets of Fig.~\ref{fig:1}), the energy loss that is pumped into the reservoir is given by
\begin{equation}
\dot E=\frac{\hbar}{4\pi}\mathrm{Tr}\left(\dot{\mathbf S}\dot{\mathbf S}^\dagger\right)=\frac{\hbar}{4\pi}\mathrm{Tr}\left(\frac{\partial\mathbf S}{\partial \mathbf n}\frac{\partial\mathbf S^\dagger}{\partial \mathbf n}\right) \dot{\mathbf n}^2\equiv D_n \dot{\mathbf n}^2.\label{eq:epump}
\end{equation}
Here we define $D_n\equiv(\hbar/4\pi)\mathrm{Tr}[(\partial\mathbf S/\partial\mathbf n)(\partial\mathbf S^\dagger/\partial\mathbf n)]$. Comparing Eqs.~\eqref{eq:edot} and \eqref{eq:epump}, we obtain
\begin{equation}
D_n=\frac{\mu_0 M_s A}{\gamma}\alpha_n L,
\label{eq:alphan}
\end{equation}
where we replace the volume $V$ by the product of the cross-sectional area $A$ and the length $L$ of the AFM. We can express $\alpha_m$ in the same manner,
\begin{equation}
D_m=\frac{\mu_0 M_s A}{\gamma}\alpha_m L
\label{eq:alpham}
\end{equation}
with $D_m\equiv(\hbar/4\pi)\mathrm{Tr}[(\partial\mathbf S/\partial\mathbf m)(\partial\mathbf S^\dagger/\partial\mathbf m)]$. Using Eqs.~\eqref{eq:alphan} and \eqref{eq:alpham}, we calculate the energy dissipation as a function of the length $L$ and extract the damping parameters $\alpha_{n(m)}$ via a linear least squares fitting. Note that the above formalism can be generalized to include noncollinear AFM, such as DWs in AFMs, by introducing the position-dependent order parameters $\mathbf n(\mathbf r)$ and $\mathbf m(\mathbf r)$. It can also be extended for the AFMs containing more than two sublattices, which may not be collinear with one another \cite{Kohn2013}. For the latter case, one has to redefine the proper order parameters instead of $\mathbf n$ and $\mathbf m$ \cite{Yuan2017}.

{\it\color{red}First-principles calculations.---}The above formalism is implemented using the first-principles scattering calculation and is applied here in studying the damping of metallic AFMs including PtMn, IrMn, PdMn and FeMn. The lattice constants and magnetic configurations are the same as in the reported first-principles calculations \cite{Zhang2014}. Here we take tetragonal PtMn as an example to illustrate the computational details. A finite thickness ($L$) of PtMn is connected to two semi-infinite Au leads along (001) direction. The lattice constant of Au is made to match that of the $a$ axis of PtMn. The electronic structures are obtained self-consistently within the density functional theory implemented with a minimal basis of the tight-binding linear muffin-tin orbitals (TB LMTOs) \cite{Andersen1986}. The magnetic moment of every Mn atom is 3.65 $\mu_B$ and Pt atoms are not magnetized.

To evaluate $\alpha_n$ and $\alpha_m$, we first construct a lateral 10$\times$10 supercell including 100 atoms per atomic layer in the scattering region, where the atoms are randomly displaced from their equilibrium lattice sites using a Gaussian distribution with the root-mean-square (RMS) displacement $\Delta$ \cite{Liu2011,Liu2015}. The value of $\Delta$ is chosen to reproduce typical experimental resistivity of the corresponding bulk AFM. The scattering matrix $\mathbf S$ are obtained using a first-principles ``wave-function matching'' scheme that is also implemented with TB LMTOs \cite{Xia2006} and its derivative is obtained by finite-difference method \cite{SM}.

Figure~\ref{fig:1}(a) shows the calculated energy pumping rate $D_n$ of PtMn as a function of $L$ for $\mathbf n$ along the $c$ axis with $\Delta/a=0.049$. The total pumping rate (solid symbols) increases linearly with increasing the volume of the AFM. A linear least squares fitting yields $\alpha_n=(0.67\pm0.02)\times10^{-3}$, as plotted by the solid line. The finite intercept of the solid line corresponding to the interface-enhanced energy dissipation, which is essentially the spin pumping effect at the AFM$|$Au interface \cite{Jia2011,Cheng2014}. The N{\'e}el order induced damping $\alpha_n$ completely results from spin-orbit coupling (SOC). If we artificially turn SOC off, the calculated pumping rate is independent of the volume of the AFM indicating $\alpha_n=0$. This is because the spin space is decoupled from the real space without SOC and the energy is then invariant with respect to the direction of $\mathbf n$. The spin pumping effect is nearly unchanged by the SOC.

The energy pumping rate $D_m$ of PtMn with $\mathbf n$ along the $c$ axis is plotted in Fig.~\ref{fig:1}(b), where we find three important features. (1) The extracted value of $\alpha_m=0.59\pm0.02$, which is nearly 1000 times larger than $\alpha_n$. (2) Turning SOC off only slightly increases the calculated $\alpha_m$ indicating that SOC is not the main dissipative mechanism of $\alpha_m$. The difference between the solid and empty circles in Fig.~\ref{fig:1}(b) can be attributed to the SOC-induced variation of electronic structure near the Fermi level. To see more clearly the different influence of SOC on $\alpha_m$ and $\alpha_n$, we plot in the inset of Fig.~\ref{fig:1}(b) the calculated damping parameters as a function of SOC strength. Indeed, as the SOC strength $\xi_{\rm SO}$ is artificially tuned from its real value to zero, $\alpha_n$ decreases dramatically and tends to vanish at $\xi_{\rm SO}=0$, while $\alpha_m$ is less sensitive to $\xi_{\rm SO}$ than $\alpha_n$. (3) The intercepts of the solid and dashed lines are both vanishingly small indicating that this specific mode does not pump spin current into the nonmagnetic leads. The pumped spin current from an AFM generally reads $\mathbf I_s^{\rm pump}\propto\mathbf n\times\dot{\mathbf n}+\mathbf m\times\dot{\mathbf m}$ \cite{Cheng2014}. For the mode depicted in Fig.~\ref{fig:1}(b), one has $\dot{\mathbf n}=0$ and $\dot{\mathbf m}\|\mathbf m$ such that $\mathbf I_s^{\rm pump}=0$.

\begin{figure}[t]
\includegraphics[width=0.85\columnwidth]{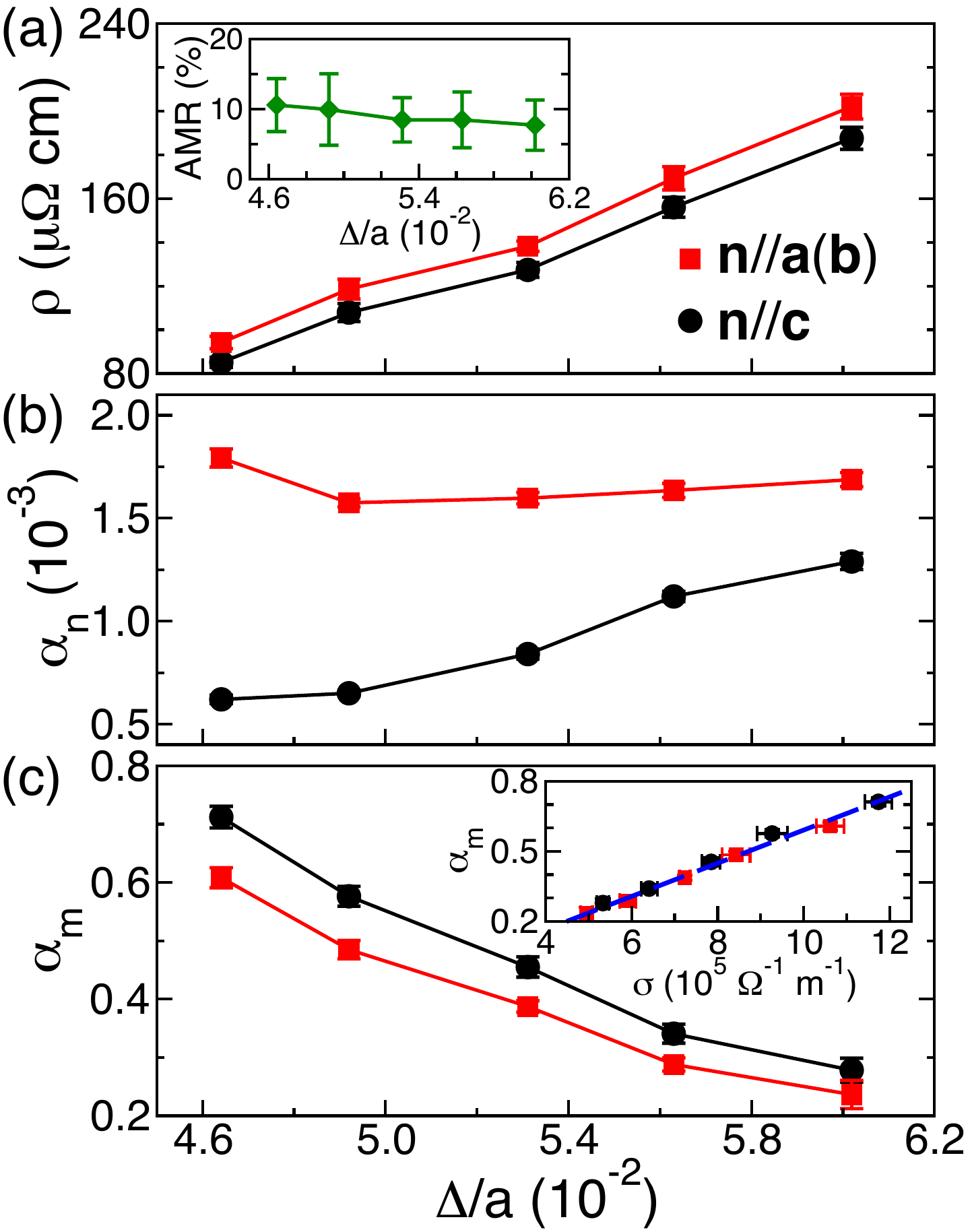}
\caption{Calculated resistivity (a) and damping parameters $\alpha_n$ (b) and $\alpha_m$ (c) of PtMn as a function of the RMS of atomic displacements. The red squares and black circles are calculated with $\mathbf n$ along $a$ axis and $c$ axis, respectively. The inset of (a) shows the calculated AMR. $\alpha_m$ is replotted as a function of conductivity in the inset of (c). The blue dashed line illustrates the linear dependence. }\label{fig:2}
\end{figure}
To explore the disorder dependence of the damping parameters $\alpha_n$ and $\alpha_m$, we further perform the calculation by varying the RMS of atomic displacements $\Delta$. Figure~\ref{fig:2}(a) shows that the calculated resistivity increases monotonically with increasing $\Delta$. The resistivity $\rho_c$ with $\mathbf n$ along $c$ axis is lower than $\rho_a$ with $\mathbf n$ along $a$ axis. The anisotropic magnetoresistance (AMR) defined by $(\rho_a-\rho_c)/\rho_c$ is about 10\%, which slightly decreases with increasing $\Delta$, as plotted in the inset of Fig.~\ref{fig:2}(a). The large AMR in PtMn is useful for experimental detection of the N{\'e}el order. The calculated AMR seems to be an order of magnitude larger than the reported values in literature \cite{Wang2012,Fina2014,Moriyama2015}. We may attribute the difference to the surface scattering in thin-film samples and other types of disorder that have been found to decrease the AMR of ferromagnetic metals and alloys \cite{McGuire1975}.

$\alpha_n$ of PtMn plotted in Fig.~\ref{fig:2}(b) is of the order of 10$^{-3}$, which is comparable with the magnitude of the Gilbert damping of ferromagnetic transition metals \cite{Heinrich1966,Bhagat1974,Heinrich1979,Liu2011}. For $\mathbf n$ along $a$ axis, $\alpha_n$ shows a weak nonmonotonic dependence on disorder, while $\alpha_n$ for $\mathbf n$ along $c$ axis increases monotonically. With the relativistic SOC, the electronic structure of an AFM depends on the orientation of $\mathbf n$. When $\mathbf n$ varies in time, the occupied energy bands may be lifted above the Fermi level. Then a longer relaxation time (weaker disorder) gives rise to a larger energy dissipation, corresponding to the increase in $\alpha_n$ with decreasing $\Delta$ at small $\Delta$. It is analogous to the intraband transitions accounting for the conductivity-like behavior of Gilbert damping at low temperature in the torque-correlation model \cite{Kambersky1976,Gilmore2007,*Gilmore2010}. Sufficiently strong disorder renders the system isotropic and the variation of $\mathbf n$ does not lead to electronic excitation but scattering of conduction electrons by disorder still dissipates energy into the lattice through SOC. The higher the scattering rate, the larger is the energy dissipation rate corresponding to the contribution of the interband transitions \cite{Kambersky1976,Gilmore2007,*Gilmore2010}. Therefore, $\alpha_n$ shares the same physical origin as the Gilbert damping of metallic FMs.

The value of $\alpha_m$ is about three orders of magnitude larger than $\alpha_n$ and it decreases monotonically with increasing the structural disorder, as shown in Fig.~\ref{fig:2}(c). This remarkable difference can be attributed to the energy involved in the dynamical motion of $\mathbf m$ and $\mathbf n$. While the precession of $\mathbf n$ only changes the magnetic anisotropy energy in an AFM, the variation of $\mathbf m$ changes the exchange energy that is in magnitude much larger than the magnetic anisotropy energy.

Physically, $\alpha_m$ can be understood in terms of spin pumping \cite{Tserkovnyak:prl02a,*Tserkovnyak:prb02b,Liu2014} between the two sublattices of an AFM. The sublattice $\mathbf m_2$ pumps a spin current that can be absorbed by $\mathbf m_1$ resulting in a damping torque exerted on $\mathbf m_1$ as $\alpha'\mathbf m_1\times[\mathbf m_1\times(\mathbf m_2\times\dot{\mathbf m}_2)]$. Here $\alpha'$ is a dimensionless parameter to describe the strength of the spin pumping. This torque can be simplified to be $\alpha'\mathbf m_1\times\dot{\mathbf m}_2$ by neglecting the high-order terms of the total magnetization $\mathbf m$. In addition, the spin pumping by $\mathbf m_1$ also contributes to the damping of the sublattice $\mathbf m_1$ that is equivalent to a torque $\alpha'\mathbf m_1\times\dot{\mathbf m}_1$ exerted on $\mathbf m_1$. Taking the inter-sublattice spin pumping into account, we are able to derive Eqs.~\eqref{eq:norder} and \eqref{eq:morder} and obtain the damping parameters $\alpha_n=\alpha_{0}/2$ and $\alpha_m=(\alpha_0+2\alpha')/2$ \cite{SM}. Here $\alpha_0$ is the intrinsic damping due to SOC for each sublattice. It is worth noting that the spin pumping strength within a metal is proportional to its conductivity \cite{Foros2008,Zhang2009,Yuan2014,*Yuan2016}. We replot $\alpha_m$ as a function of conductivity in the inset of Fig.~\ref{fig:2}(c), where a general linear dependence is seen for both $\mathbf n$ along $a$ axis and $c$ axis.

We list in Table~\ref{tab:1} the calculated $\rho$, $\alpha_n$ and $\alpha_m$ for typical metallic AFMs including PtMn, IrMn, PdMn and FeMn. For IrMn, $\alpha_m$ is only 10 times larger than $\alpha_n$, while $\alpha_m$ of the other three materials are about three orders of magnitude larger than their $\alpha_n$.

\begin{table}[b]
\centering
\caption{Calculated resistivity and damping parameters for the N{\'e}el order $\mathbf n$ along $a$ axis and $c$ axis.}
\begin{ruledtabular}
\begin{tabular}{lcccc}
            \multicolumn{1}{c} {AFM} &
            \multicolumn{1}{c} {$\mathbf n$} &
            \multicolumn{1}{c} {$\rho$ ($\mu\Omega$ cm)} &
            \multicolumn{1}{c} {$\alpha_n$ ($10^{-3}$) } &
            \multicolumn{1}{c} {$\alpha_m$} \\
\hline
 PtMn  & $a$ axis & 119$\pm$5 & 1.60$\pm$0.02 & 0.49$\pm$0.02 \\
           & $c$ axis & 108$\pm$4 & 0.67$\pm$0.02 & 0.59$\pm$0.02 \\
 IrMn   & $a$ axis & 116$\pm$2 & 10.5$\pm$0.2 & 0.10$\pm$0.01 \\
           & $c$ axis & 116$\pm$2 & 10.2$\pm$0.3 & 0.10$\pm$0.01 \\
 PdMn & $a$ axis & 120$\pm$8 & 0.16$\pm$0.02 & 1.1$\pm$0.10 \\
           & $c$ axis & 121$\pm$8 & 1.30$\pm$0.10 & 1.30$\pm$0.10 \\
 FeMn & $a$ axis & 90$\pm$1 & 0.76$\pm$0.04 & 0.38$\pm$0.01 \\
           & $c$ axis & 91$\pm$1 & 0.82$\pm$0.03 & 0.38$\pm$0.01 \\
\end{tabular}
\end{ruledtabular}
\label{tab:1}
\end{table}

\begin{figure}[t]
\includegraphics[width=\columnwidth]{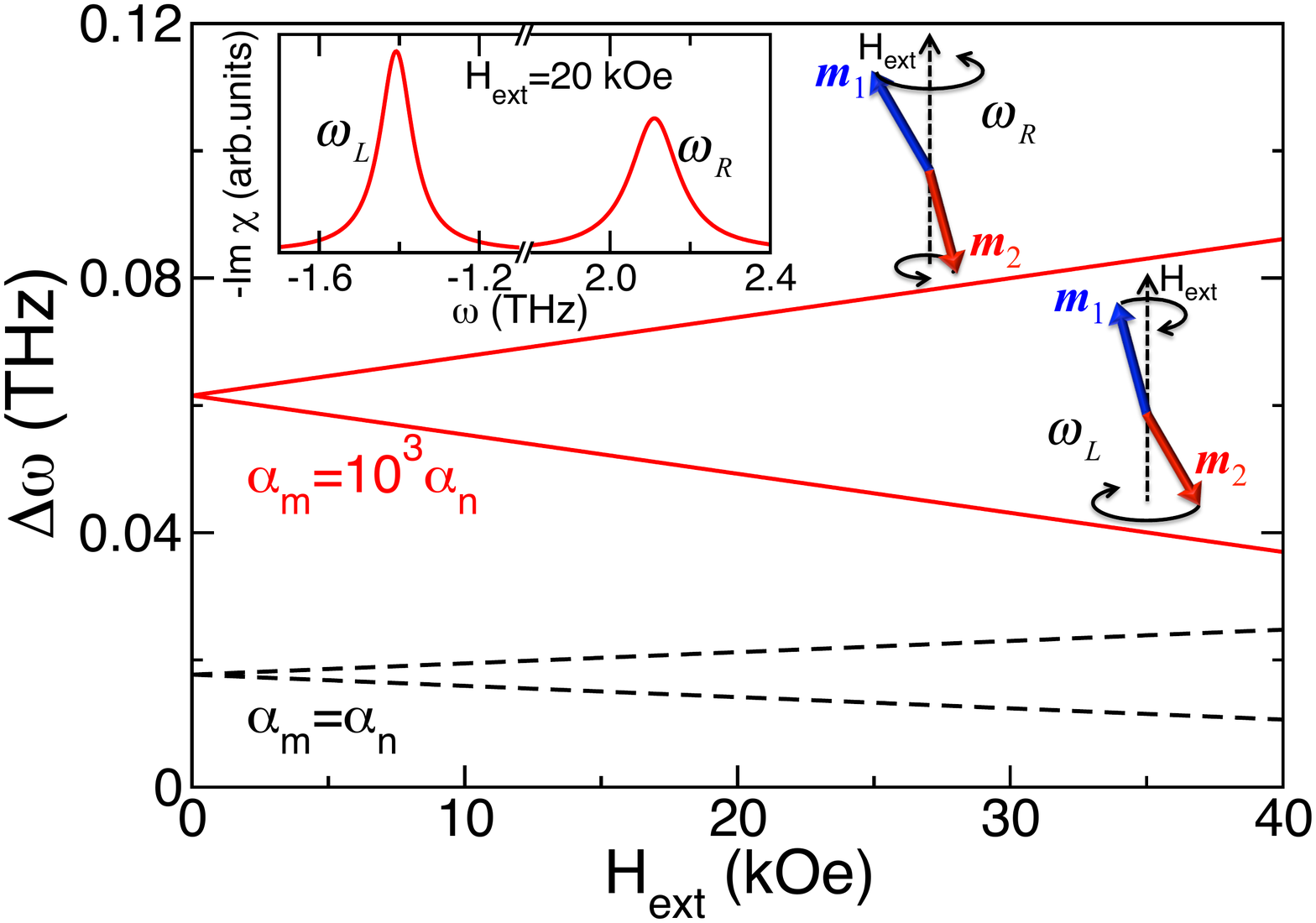}
\caption{Linewidth of AFMR as a function of the external magnetic field. The black dashed lines and red solid lines are calculated with $\alpha_m=\alpha_n$ and $\alpha_m=10^3\alpha_n$, respectively. Inset: the imaginary part of susceptibility as a function of the frequency for the external magnetic field $H_{\rm ext}=20$ kOe and $\alpha_m=10^3\alpha_n$. The cartoons illustrate the corresponding dynamical modes. Here we use $H_E=10^3$ kOe, $H_A=5$~kOe and $\alpha_n=0.001$.}\label{fig:3}
\end{figure}
{\it\color{red}Antiferromagnetic resonance.---}
Keffer and Kittel formulated antiferromagnetic resonance (AFMR) without damping \cite{Keffer1952} and determined the resonant frequencies that depend on the external field $H_{\rm ext}$, exchange field $H_E$ and anisotropy field $H_A$, $\omega_{\rm res}=\gamma\left[H_{\rm ext}\pm\sqrt{H_A(2H_E+H_A)}\right]$. Here we follow their approach, in which $H_{\rm ext}$ is applied along the easy axis and the transverse components of $\mathbf m_1$ and $\mathbf m_2$ are supposed to be small. Taking both the intrinsic damping due to SOC and spin pumping between the two sublattices into account, we solve the dynamical equations of AFMR and find the frequency-dependent susceptibility $\bm{\chi}(\omega)$ that is defined by $\mathbf n_{\perp}(\omega)=\bm{\chi}(\omega)\cdot\mathbf h_{\perp}(\omega)$. Here $\mathbf n_{\perp}$ and $\mathbf h_{\perp}$ are the transverse components of the N{\'e}el order and microwave field, respectively. The imaginary part of the diagonal element of $\bm{\chi}(\omega)$ with $H_{\rm ext}=20$ kOe is plotted in the inset of Fig.~\ref{fig:3}, where two resonance modes can be identified. The precessional modes for the positive ($\omega_R$) and negative frequency ($\omega_L$) are schematically depicted in Fig.~\ref{fig:3}. The linewidth of the AFMR $\Delta\omega$ can be determined from the imaginary part of the (complex) eigen-frequency \cite{Saib2004} by solving $\det\vert\bm{\chi}^{-1}(\omega)\vert=0$ and is plotted in Fig.~\ref{fig:3} as a function of $H_{\rm ext}$. Without $H_{\rm ext}$, the two modes have the same linewidth. A finite external field increases the linewidth of $\omega_R$ and decreases that of $\omega_L$, both linearly. By including the spin pumping between two sublattices, both the linewidth at $H_{\rm ext}=0$ and the slope of $\Delta\omega$ as a function of $H_{\rm ext}$ increase by a factor of about 3.5. It indicates that the spin pumping effect between the two sublattices plays an important role in the magnetization dynamics of metallic AFMs.

{\it\color{red}Conclusions.---}We have generalized the scattering theory of magnetization dissipation in FMs to be applicable for AFMs. Using first-principles scattering calculation, we find the damping parameter accompanying the motion of magnetization ($\alpha_m$) is generally much larger than that associated with the motion of the N{\'e}el order ($\alpha_n$) in metallic AFMs PtMn, IrMn, PdMn and FeMn. While $\alpha_n$ arises from the spin-orbit interaction, $\alpha_m$ is mainly contributed by the spin pumping between the two sublattices in an AFM via exchange interaction. Taking AFMR as an example, we demonstrate that the linewidth can be significantly enhanced by the giant value of $\alpha_m$. Our findings suggest that the magnetization dynamics of AFMs shall be revisited with the damping effect properly included.

\begin{acknowledgments}
We would like to thank the helpful discussions with X. R. Wang. This work was financially supported by the National Key Research and Development Program of China (2017YFA0303300) and National Natural Science Foundation of China (Grants No. 61774018, No. 61704071, No. 11734004, No. 61774017 and No. 21421003).
\end{acknowledgments}

\pagebreak
\begin{widetext}
\clearpage
\begin{center}
\textbf{\large Supplementary Material for ``Mode-Dependent Damping in Metallic Antiferromagnets Due to Inter-Sublattice Spin Pumping'' }

\vspace{3mm}
Qian Liu,$^{1,\ast}$ H. Y. Yuan,$^{2,\ast}$ Ke Xia,$^{1,3}$ and Zhe Yuan$^{1,\dagger}$

\vspace{2mm}
$^1${\small\it The Center for Advanced Quantum Studies and Department of Physics, Beijing Normal University, Beijing 100875, China}

$^2${\small\it Department of Physics, Southern University of Science and Technology of China, Shenzhen, Guangdong, 518055, China}

$^3${\small\it Synergetic Innovation Center for Quantum Effects and Applications (SICQEA), Hunan Normal University, Changsha 410081, China}

\end{center}
\setcounter{equation}{0}
\setcounter{figure}{0}
\setcounter{table}{0}
\setcounter{page}{1}
\makeatletter
\renewcommand{\theequation}{S\arabic{equation}}
\renewcommand{\thefigure}{S\arabic{figure}}
\renewcommand{\thetable}{S\Roman{figure}}
\renewcommand{\bibnumfmt}[1]{[S#1]}
\renewcommand{\citenumfont}[1]{S#1}

{\small In the Supplemental Material, we present the detailed derivation of the energy pumping arising from antiferromagnetic dynamics, the implementation of calculating the derivatives of scattering matrix and derivation of dynamic equations of $\mathbf n$ and $\mathbf m$ including the spin pumping between sublattices.}

\section{Derivation of energy dissipation in antiferromagnetic dynamics}
We consider a collinear antiferromagnet (AFM) with two sublattices, both of which have the magnetization $M_s$. The magnetization directions are denoted by the unit vectors $\bm{m}_1$ and
$\bm{m}_2$. Then we are able to define the total magnetization $\mathbf m=\mathbf m_1+\mathbf m_2$ and the N{\'e}el order parameter $\mathbf n=\mathbf m_1-\mathbf m_2$. The dynamic equations of $\bm{m}$ and $\bm{n}$ can be written as \cite{Hals2011s,Gomonay2014s}
\begin{eqnarray}
\dot{\bm{m}}&=&-\gamma(\bm{m}\times\bm{h}_m+\bm{n}\times\bm{h}_n)
             +\alpha_m\bm{m}\times\dot{\bm{m}}+\alpha_n\bm{n}\times\dot{\bm{n}},\label{eq:mfinal}\\
\dot{\bm{n}}&=&-\gamma(\bm{m}\times\bm{h}_n+\bm{n}\times\bm{h}_m)
             +\alpha_m\bm{n}\times\dot{\bm{m}}+\alpha_n\bm{m}\times\dot{\bm{n}}.
\label{eq:nfinal}
\end{eqnarray}
Here $\mathbf h_m$ and $\mathbf h_n$ are the effective fields acting on the total magnetization and the N{\'e}el order. Specifically, if the free energy is written as $\mathcal F=\mu_0 M_s V \mathcal E$, where $\mu_0$ is the vacuum permeability, $V$ is the volume of the AFM, and $\mathcal E$ is a reduced free energy density, one has \cite{Hals2011s}
\begin{equation}
\mathbf h_m=-\frac{\delta\mathcal E}{\delta\mathbf m},~\mathrm{and}\,\,\,\mathbf h_n=-\frac{\delta\mathcal E}{\delta\mathbf n}.
\end{equation}
In Eqs.~\eqref{eq:mfinal} and \eqref{eq:nfinal}, $\alpha_m$ and $\alpha_n$ are used to characterize the damping due to the variation of the magnetization and the N{\'e}el order, respectively.

If $\mathbf m$ and $\mathbf n$ are the only time-varying parameters in the system, the energy dissipation can be represented by
\begin{eqnarray}
\dot{E}&=&-\dot{\mathcal F}=-\mu_0 M_s V\dot{\mathcal E}\nonumber\\
&=&\mu_0 M_s V \left[\dot{\mathbf m}\cdot\left(-\frac{\delta \mathcal E}{\delta\mathbf m}\right)+\dot{\mathbf n}\cdot\left(-\frac{\delta\mathcal E}{\delta \mathbf n}\right)\right]\nonumber\\
&=&\mu_0 M_s V\left(\dot{\mathbf m}\cdot\mathbf h_m+\dot{\mathbf n}\cdot\mathbf h_n\right).\label{eq:diss}
\end{eqnarray}
We then insert the dynamic Eqs.~\eqref{eq:mfinal} and \eqref{eq:nfinal} into the above Eq.~\eqref{eq:diss} and obtain
\begin{eqnarray}
\frac{\dot{E}}{\mu_0 M_s V}&=&\left[-\gamma\left(\mathbf m\times\mathbf h_m+\mathbf{n}\times\mathbf{h}_n\right)
             +\alpha_m\mathbf{m}\times\dot{\mathbf{m}}+\alpha_n\mathbf{n}\times\dot{\mathbf{n}}\right]\cdot\mathbf{h}_m\nonumber\\
             &&+[-\gamma\left(\mathbf{m}\times\mathbf{h}_n+\mathbf{n}\times\mathbf{h}_m\right)
             +\alpha_m\mathbf{n}\times\dot{\mathbf{m}}+\alpha_n\mathbf{m}\times\dot{\mathbf{n}}]\cdot\mathbf{h}_n\nonumber\\
&=&-\gamma\mathbf{n}\times\mathbf{h}_n\cdot\mathbf{h}_m+\left(\alpha_m\mathbf{m}\times\dot{\mathbf{m}}+\alpha_n\mathbf{n}\times\dot{\mathbf{n}}\right)\cdot\mathbf{h}_m-\gamma\mathbf{n}\times\mathbf{h}_m\cdot\mathbf{h}_n+\left(\alpha_m\mathbf{n}\times\dot{\mathbf{m}}+\alpha_n\mathbf{m}\times\dot{\mathbf{n}}\right)\cdot\mathbf{h}_n\nonumber\\
&=&\left(\alpha_m\mathbf{m}\times\dot{\mathbf{m}}+\alpha_n\mathbf{n}\times\dot{\mathbf{n}}\right)\cdot\mathbf{h}_m+\left(\alpha_m\mathbf{n}\times\dot{\mathbf{m}}+\alpha_n\mathbf{m}\times\dot{\mathbf{n}}\right)\cdot\mathbf{h}_n\nonumber\\
&=&\alpha_m\left(\mathbf{m}\times\dot{\mathbf{m}}\cdot\mathbf{h}_m + \mathbf{n}\times\dot{\mathbf{m}}\cdot\mathbf{h}_n\right)+\alpha_n\left(\mathbf{n}\times\dot{\mathbf{n}}\cdot \mathbf{h}_m + \mathbf{m}\times\dot{\mathbf{n}}\cdot \mathbf{h}_n\right)\nonumber\\
&=&\alpha_m\left(\mathbf{h}_m\times\mathbf{m}\cdot\dot{\mathbf{m}}+\mathbf{h}_n\times\mathbf{n}\cdot\dot{\mathbf{m}}\right)+\alpha_n\left(\mathbf{h}_m\times\mathbf{n}\cdot\dot{\mathbf{n}}+\mathbf{h}_n\times\mathbf{m}\cdot\dot{\mathbf{n}}\right)\nonumber\\
&=&\alpha_m\left(\mathbf{h}_m\times\mathbf{m}+\mathbf{h}_n\times\mathbf{n}\right)\cdot\dot{\mathbf{m}}+\alpha_n\left(\mathbf{h}_m\times\mathbf{n}+\mathbf{h}_n\times\mathbf{m}\right)\cdot\dot{\mathbf{n}}\nonumber\\
&=&\alpha_m\frac{1}{\gamma}\left(\dot{\mathbf{m}}-\alpha_m\mathbf{m}\times\dot{\mathbf{m}}-\alpha_n\mathbf{n}\times\dot{\mathbf{n}}\right)\cdot\dot{\mathbf{m}}+\alpha_n\frac{1}{\gamma}\left(\dot{\mathbf{n}}-\alpha_m\mathbf{n}\times\dot{\mathbf{m}}-\alpha_n\mathbf{m}\times\dot{\mathbf{n}}\right)\cdot\dot{\mathbf{n}}\nonumber\\
&=&\frac{\alpha_m}{\gamma}\dot{\mathbf{m}}^2-\frac{\alpha_m \alpha_n}{\gamma}\mathbf{n}\times\dot{\mathbf{n}}\cdot\dot{\mathbf{m}}+\frac{\alpha_n}{\gamma}\dot{\mathbf{n}}^2-\frac{\alpha_n \alpha_m}{\gamma}\mathbf{n}\times\dot{\mathbf{m}}\cdot\dot{\mathbf{n}}\nonumber\\
&=&\frac{1}{\gamma}\left(\alpha_m\dot{\mathbf{m}}^2+\alpha_n\dot{\mathbf{n}}^2\right).
\end{eqnarray}
Therefore the energy dissipation during antiferromagnetic dynamics can be eventually obtained 
\begin{equation}
\dot{E}=\frac{\mu_0 M_s V}{\gamma}\left(\alpha_m\dot{\mathbf{m}}^2+\alpha_n\dot{\mathbf{n}}^2\right).\label{eq:dissfinal}
\end{equation}

\begin{figure}
\begin{center}
\includegraphics[width=0.5\columnwidth]{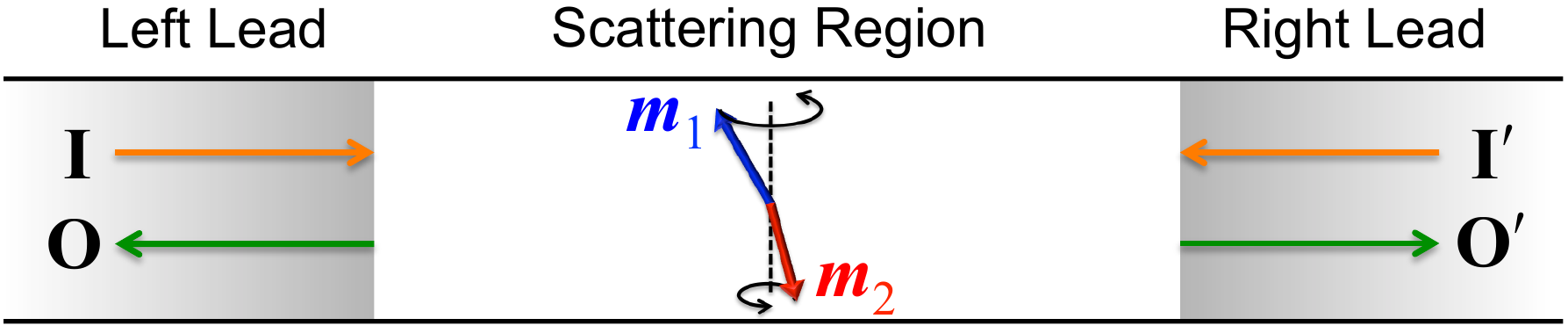}
\caption{Schematic illustration of the scattering geometry that is used in the first-principles calculations. Since both the left and right leads are semi-infinite with periodic crystalline structure, the propagating (incoming and outgoing) Bloch states can be obtained by solving the Kohn-Sham equation self-consistently. Then the transmission and reflection coefficients can be solved using the numerical technique called ``wave function matching'' \cite{Xia2006s}.}
\label{fig1}
\end{center}
\end{figure}

\section{Calculating the derivative of scattering matrix}
Noting that the energy dissipation in a scattering geometry, i.e. the left lead--scattering region--the right lead (see Fig.~\ref{fig1}), can be written in terms of the parametric pumping~\cite{Avron2001s}
\begin{equation}
\dot{E}=\frac{\hbar}{4\pi}\mathrm{Tr}\left(\dot{\mathbf S}\dot{\mathbf S}^{\dagger}\right).\label{eq:edot}
\end{equation}
Here $\mathbf S$ is the scattering matrix. Supposing only the magnetic order $\bm{\zeta}$ ($\bm{\zeta}=\mathbf m$ or $\bm{\zeta}=\mathbf n$) of the system is varying in time, one can rewrite Eq.~\eqref{eq:edot} as
\begin{equation}
\dot{E}=\frac{\hbar}{4\pi}\mathrm{Tr}\left(\frac{\partial\mathbf S}{\partial\bm{\zeta}}\frac{\partial\mathbf S^{\dagger}}{\partial\bm{\zeta}}\right)\dot{\bm{\zeta}}^2\equiv\mathbf D_{\zeta}\dot{\bm{\zeta}}^2.\label{eq:edot2}
\end{equation}
The quantity $\mathbf D_{\bm{\zeta}}$ is generally a positive-definite and symmetric tensor \cite{Brataas2011s} with its elements defined by
\begin{equation}
D_{\zeta}^{ij}=\frac{\hbar}{4\pi}\mathrm{Tr}\left(\frac{\partial\mathbf S}{\partial\zeta_i}\frac{\partial\mathbf S^{\dagger}}{\partial\zeta_j}\right).\label{eq:dzeta}
\end{equation}
Noting that $\partial\mathbf S/\partial\zeta_i$, $\partial\mathbf S^{\dagger}/\partial\zeta_j$ and their product are all matrices, so we rewrite Eq.~\eqref{eq:dzeta} in terms of the specific matrix elements as
\begin{eqnarray}
D_{\zeta}^{ij}=\frac{\hbar}{4\pi}\sum_{\mu}\left(\frac{\partial\mathbf S}{\partial\zeta_i}\frac{\partial\mathbf S^{\dagger}}{\partial\zeta_j}\right)_{\mu\mu}
=\frac{\hbar}{4\pi}\sum_{\mu}\sum_{\nu}\frac{\partial S_{\mu\nu}}{\partial\zeta_i}\frac{\partial \left(\mathbf S^\dagger\right)_{\nu\mu}}{\partial\zeta_j}
=\frac{\hbar}{4\pi}\sum_{\mu}\sum_{\nu}\frac{\partial S_{\mu\nu}}{\partial\zeta_i}\left(\frac{\partial S_{\mu\nu}}{\partial\zeta_j}\right)^\ast.
\end{eqnarray}
In particular, for $i=j$, we have the diagonal elements of $\mathbf D_\zeta$
\begin{equation}
D_{\zeta}^{ii}=\frac{\hbar}{4\pi}\sum_{\mu,\nu}\left\vert\frac{\partial S_{\mu\nu}}{\partial\zeta_i}\right\vert^2,
\end{equation}
which is a real number. All the remaining task is to numerically calculate the derivatives of the scattering matrix elements $\partial S_{\mu\nu}/\partial \zeta_i$.

In the following, we take $\bm{\zeta}=\mathbf n$ as an example and illustrate the calculation of $\partial S_{\mu\nu}/\partial \zeta_i$. Considering the N{\'e}el order along $z$-axis, i.e. $\mathbf m_1=-\mathbf m_2=\hat z$ and $\mathbf n=2\hat z$, one can calculate the scattering matrix $\mathbf S(\mathbf n)$. Then we add an infinitesimal transverse component $\Delta\mathbf n=\eta \hat x$ onto the N{\'e}el order so that the new N{\'e}el order becomes $\mathbf n'=2\hat z+\eta\hat x$. (In practice, we find that the calculated results are well converged with $\eta$ in the range of $10^{-3}$--$10^{-5}$.) Under such a magnetic configuration, we redo the scattering calculation to find another scattering matrix $\mathbf S(\mathbf n')$. The derivatives of the matrix element $S_{\mu\nu}$ can be obtained by
\begin{equation}
\frac{\partial S_{\mu\nu}}{\partial n_x}=\frac{S'_{\mu\nu}-S_{\mu\nu}}{\eta}.
\end{equation} 
In the same manner, we can find another scattering matrix $\mathbf S''$ at $\mathbf n''=2\hat z+\eta\hat y$ and consequently we have
\begin{equation}
\frac{\partial S_{\mu\nu}}{\partial n_y}=\frac{S''_{\mu\nu}-S_{\mu\nu}}{\eta}.
\end{equation} 
Finally, we find that the calculated off-diagonal elements $D^{xy}_{n(m)}$ and $D^{yx}_{n(m)}$ are much smaller than the diagonal elements $D^{xx}_{n(m)}$ and $D^{yy}_{n(m)}$. The latter two are nearly the same. So we take their average in practice, i.e. $D_n=\left(D_n^{xx}+D_n^{yy}\right)/2$ and $D_m=\left(D_m^{xx}+D_m^{yy}\right)/2$.

\section{Dynamical equations with inter-sublattice spin pumping}
We start from the coupled dynamical equations of an AFM with the sublattice index $i=1,2$,
\begin{equation}
\dot{\mathbf m}_i=-\gamma\mathbf m_i\times\mathbf h_i+\alpha_0\mathbf m_i\times\dot{\mathbf m}_i.
\end{equation}
Here $\mathbf h_i$ is the effective field exerted on $\mathbf m_i$, which can be calculated from the functional derivative of the free energy $\mathcal F$ as
\begin{equation}
\mathbf h_i=-\frac{1}{\mu_0M_sV}\frac{\delta\mathcal F}{\delta m_i}.
\end{equation}
$\alpha_0$ is the damping parameter, which must be equal for $\mathbf m_1$ and $\mathbf m_2$ because of the permutation symmetry. Now we consider the spin pumping effect that discussed in the main text. The spin pumping by the sublattice $\mathbf m_1$ contributes a dissipative torque $\alpha'\mathbf m_1\times\dot{\mathbf m}_1$ that is exerted on $\mathbf m_1$. Here $\alpha'$ is a dimensionless parameter to quantify the magnitude of the inter-sublattice spin pumping. The pumped spin current by $\mathbf m_1$ can be absorbed by $\mathbf m_2$ resulting in a damping-like torque $\mathbf m_2\times\left[\mathbf m_2\times\left(\alpha'\mathbf m_1\times\dot{\mathbf m}_1\right)\right]\approx\alpha'\mathbf m_2\times\dot{\mathbf m}_1$, which is exerted on $\mathbf m_2$. In the same manner, we can identify two torques due to the spin pumping of $\mathbf m_2$: $\alpha'\mathbf m_1\times\dot{\mathbf m}_2$ exerted on $\mathbf m_1$ and $\alpha'\mathbf m_2\times\dot{\mathbf m}_2$ exerted on $\mathbf m_2$. Eventually, we obtain the coupled dynamical equations by including the inter-sublattice spin pumping as
\begin{eqnarray}
&&\dot{\mathbf m}_1=-\gamma\mathbf m_1\times\mathbf h_1+\left(\alpha_0+\alpha'\right)\mathbf m_1\times\dot{\mathbf m}_1+\alpha'\mathbf m_1\times\dot{\mathbf m}_2,\nonumber\\
&&\dot{\mathbf m}_2=-\gamma\mathbf m_2\times\mathbf h_2+\left(\alpha_0+\alpha'\right)\mathbf m_2\times\dot{\mathbf m}_2+\alpha'\mathbf m_2\times\dot{\mathbf m}_1.\label{eq:coupledmfull}
\end{eqnarray}
The above form of the dynamical equations can be rigorously derived using the Rayleigh functional to describe the dissipation \cite{Yuan2017s}.

In the following, we rewrite Eq.~\eqref{eq:coupledmfull} into the dynamical equations of the total magnetization $\mathbf m=\mathbf m_1+\mathbf m_2$ and the N{\'e}el order $\mathbf n=\mathbf m_1-\mathbf m_2$. The effective field $\mathbf h_i$ can be transformed as
\begin{eqnarray}
&&\mathbf h_1=-\frac{1}{\mu_0M_sV}\frac{\delta\mathcal F}{\delta\mathbf m_1}=-\frac{1}{\mu_0M_sV}\left(\frac{\delta\mathcal F}{\delta\mathbf m}\frac{\partial \mathbf m}{\partial \mathbf m_1}+\frac{\delta\mathcal F}{\delta\mathbf n}\frac{\partial \mathbf n}{\partial \mathbf m_1}\right)=\mathbf h_m+\mathbf h_n,\nonumber\\
&&\mathbf h_2=-\frac{1}{\mu_0M_sV}\frac{\delta\mathcal F}{\delta\mathbf m_2}=-\frac{1}{\mu_0M_sV}\left(\frac{\delta\mathcal F}{\delta\mathbf m}\frac{\partial \mathbf m}{\partial \mathbf m_2}+\frac{\delta\mathcal F}{\delta\mathbf n}\frac{\partial \mathbf n}{\partial \mathbf m_2}\right)=\mathbf h_m-\mathbf h_n,\label{eq:hmhn}
\end{eqnarray}
where we have defined 
\begin{eqnarray}
\mathbf h_m&=&-\frac{1}{\mu_0M_sV}\frac{\delta\mathcal F}{\delta m},\nonumber\\
\mathbf h_n&=&-\frac{1}{\mu_0M_sV}\frac{\delta\mathcal F}{\delta n}.
\end{eqnarray}
Then we find
\begin{eqnarray}
\dot{\mathbf m}=\dot{\mathbf m}_1+\dot{\mathbf m}_2=-\gamma\left(\mathbf m_1\times\mathbf h_1+\mathbf m_2\times\mathbf h_2\right)+\left(\alpha_0+\alpha'\right)\left(\mathbf m_1\times\dot{\mathbf m}_1+\mathbf m_2\times\dot{\mathbf m}_2\right)+\alpha'\left(\mathbf m_1\times\dot{\mathbf m}_2+\mathbf m_2\times\dot{\mathbf m}_1\right).\label{eq:simplify1}
\end{eqnarray}
Using Eq.~\eqref{eq:hmhn}, the first term in the right-hand side of Eq.~\eqref{eq:simplify1} can be simplified as
\begin{eqnarray}
-\gamma\left(\mathbf m_1\times\mathbf h_1+\mathbf m_2\times\mathbf h_2\right)&=&-\gamma\left[\frac{\mathbf m+\mathbf n}{2}\times\left(\mathbf h_m+\mathbf h_n\right)+\frac{\mathbf m-\mathbf n}{2}\times\left(\mathbf h_m-\mathbf h_n\right)\right]=-\gamma\left(\mathbf m\times\mathbf h_m+\mathbf n\times\mathbf h_n\right).
\end{eqnarray}
The second and the third terms in the right-hand side of Eq.~\eqref{eq:simplify1} can be simplified, respectively, as
\begin{equation}
\left(\alpha_0+\alpha'\right)\left(\mathbf m_1\times\dot{\mathbf m}_1+\mathbf m_2\times\dot{\mathbf m}_2\right)=\left(\alpha_0+\alpha'\right)\left(\frac{\mathbf m+\mathbf n}{2}\times\frac{\dot{\mathbf m}+\dot{\mathbf n}}{2}+\frac{\mathbf m-\mathbf n}{2}\times\frac{\dot{\mathbf m}-\dot{\mathbf n}}{2}\right)=\frac{\alpha_0+\alpha'}{2}\left(\mathbf m\times\dot{\mathbf m}+\mathbf n\times\dot{\mathbf n}\right),
\end{equation}
and
\begin{equation}
\alpha'\left(\mathbf m_1\times\dot{\mathbf m}_2+\mathbf m_2\times\dot{\mathbf m}_1\right)=\alpha'\left(\frac{\mathbf m+\mathbf n}{2}\times\frac{\dot{\mathbf m}-\dot{\mathbf n}}{2}+\frac{\mathbf m-\mathbf n}{2}\times\frac{\dot{\mathbf m}+\dot{\mathbf n}}{2}\right)=\alpha'\left(\mathbf m\times\dot{\mathbf m}-\mathbf n\times\dot{\mathbf n}\right).
\end{equation}
Finally, Eq.~\eqref{eq:simplify1} is rewritten as
\begin{equation}
\dot{\mathbf m}=-\gamma\left(\mathbf m\times\mathbf h_m+\mathbf n\times\mathbf h_n\right)+\left(\frac{\alpha_0}{2}+\alpha'\right)\mathbf m\times\dot{\mathbf m}+\frac{\alpha_0}{2}\mathbf n\times\dot{\mathbf n}.\label{eq:simplify2}
\end{equation}
The dynamical equation of the N{\'e}el order $\mathbf n$ can be obtained in the same way
\begin{equation}
\dot{\mathbf n}=-\gamma\left(\mathbf m\times\mathbf h_n+\mathbf n\times\mathbf h_m\right)+\left(\frac{\alpha_0}{2}+\alpha'\right)\mathbf n\times\dot{\mathbf m}+\frac{\alpha_0}{2}\mathbf m\times\dot{\mathbf n}.\label{eq:simplify3}
\end{equation}
Comparing Eqs.~\eqref{eq:simplify2} and \eqref{eq:simplify3} with Eqs.~\eqref{eq:mfinal} and \eqref{eq:nfinal}, we can identify the relations of the damping parameters, i.e.
\begin{equation}
\alpha_m=\frac{\alpha_0}{2}+\alpha',\,\,\,\mathrm{and}\,\,\,\alpha_n=\frac{\alpha_0}{2}.
\end{equation}
The above relations naturally show the spin pumping effect and is consistent with our first-principles calculations.

\end{widetext}

\end{document}